\documentstyle[aps,eqsecnum,prbbib]{revtex}
\begin{document}
\draft
\title{Superconducting gap node spectroscopy using nonlinear 
electrodynamics}
\author{Igor \v{Z}uti\'c\cite{igor}  and Oriol T. Valls}
\address{School of Physics and Astronomy and Minnesota Supercomputer Institute
\\ University of Minnesota \\
Minneapolis, Minnesota 55455-0149}
\maketitle
\begin{abstract}
We present a method to  determine the nodal structure of 
 the energy gap of unconventional superconductors such as high $T_c$ 
materials.
We show how nonlinear electrodynamics phenomena in the Meissner 
regime, arising from the presence of lines on the Fermi surface where 
the superconducting energy gap is very small or zero, can be 
used to perform ``node spectroscopy'', that is, as a sensitive bulk 
probe to locate the angular position of those lines.
In calculating the nonlinear supercurrent response, we include the 
effects of orthorhombic distortion and $a-b$ plane anisotropy. 
Analytic results presented 
 demonstrate a systematic way to experimentally distinguish order 
parameters of different symmetries, including cases with mixed symmetry 
(for example, $d+s$ and $s+id$).
We consider, as suggested by various experiments, order parameters with  
predominantly $d$-wave character, and describe how to determine     
the possible  presence of other symmetries. The nonlinear      
magnetic moment displays a distinct behavior if nodes in the  gap are 
absent but regions with small, finite, values of the energy gap exist.
\end{abstract}

\section{Introduction}
In recent years a very significant effort, both experimentally and 
theoretically, has been made to determine the symmetry of the 
pairing state\cite{agl,agl2,dj,kirt,dynes} in high temperature 
superconductors (HTSC's). It is usually   
held that such an identification would provide
significant clues towards a better understanding of the 
microscopic mechanism responsible for superconductivity.  The symmetry 
properties of the pairing state itself, being connected to those of the 
superconducting energy gap, have important physical and 
technological implications. 

There is a large body of experimental results, supported by 
theoretical work, that is generally interpreted as indicative of a
pairing state at least predominantly of the $d$-wave type 
\cite{agl,agl2,dj,thermal}, (although there are 
experiments\cite{sun,chau,buan} which are 
difficult to interpret in that framework.) By this, it is meant a pairing state
which has an order parameter (OP) which vanishes, or nearly so, along
four lines on the Fermi surface (FS). The angle between such lines of nodes
(or ``quasinodes'') is presumed to be near $\pi/2$, but, because of the
orthorhombic symmetry of HTSC materials (as exemplified by the strong
anisotropy of superconducting properties\cite{ybco}
in the $a-b$ plane) this is unlikely to
be precisely correct.
Pairing states of different 
tetragonal symmetry, such as $d$- or $s$-wave, could  mix. 
Several 
proposals, mostly incorporating a predominantly $d$-wave OP,
have been set forth to describe this 
mixing\cite{mul,kim,otv,maki}. Some of them  include OP's of different 
symmetry near the surface than in the bulk\cite{bahc}. 
Many of the best probes
of the OP give information  only on the pairing state within a few
correlation lengths from the surface, which might differ from the bulk state.
Thus, large uncertainties exist.

The aim of this paper is to show that nonlinear
electromagnetic phenomena\cite{ysprl,sv,ys,zv,jcp} can provide a
systematic way to distinguish 
OP's of  different symmetries, or of mixed symmetry. We will show how
the nonlinear electrodynamic  
effects in the Meissner regime, which are due to the presence of
nodes or small minima in the 
superconducting gap, can be used to perform ``node spectroscopy'',
that is, to determine
the angular position of any lines on the FS where the absolute value of the
OP is very small or zero, and the value of the OP at these lines.
Further, the bulk character of these effects (they extend 
over a length on  the  order of the penetration depth rather than the much  
smaller coherence length) means that these determinations would refer to
the bulk, and can become very useful in avoiding material 
imperfections near the surface and other difficulties of the surface probes.

We concentrate in this work
on the nonlinear magnetic moment as a sensitive probe 
for the superconducting gap symmetry. 
In particular, we examine its transverse component (perpendicular to 
the applied field, which lies in the $a-b$ plane) 
which is determined by the quasiparticle excitations.
Other quantities, such as the penetration depth, might also  be used for
the same purpose, and will be referred to briefly. The transverse
moment, however, can be readily measured either directly or through the
torque it produces, and, as shown in Ref. \onlinecite{zv}, its size for typical
samples is large enough to be detected. 
At  temperatures much below the critical temperature, the 
excitations responsible for the nonlinear Meissner effect
are energetically favorable only 
near the nodes or quasinodes of the FS. The spectroscopy method 
described in this paper,  provides an accurate positioning of the nodes 
in the energy gap and not just information about their presence. 
Similarly we show that even in the 
absence of nodes, the presence of ``quasinodes''  would lead to several  
distinctive features in the angular and field dependence of 
the nonlinear magnetic moment. 

We will examine various pairing states that are compatible with
the crystal symmetry, including leading candidates for the pairing
state of HTSC materials, and excluding only the pure $d$-wave state which
we have extensively discussed elsewhere\cite{zv}.
In our notation for the pairing states,
we shall follow the 
discussion from Refs. \onlinecite{agl,agl2}.  We start with the spin  
singlet pairing states in a square $CuO_2$ lattice as
given in Table \ref{table1}. 
For the $YBCO$ family of cuprates, the
orthorhombic crystal distortion amounts to 
a simple change in the length of the $a$ and $b$ crystallographic constants. 
Therefore\cite{agl} the admissible mixing 
of OP's include combinations of $d_{x^2-y^2}$ and $s$, or of  
$s^-$ and $d_{xy}$ symmetries only. The nodes of the gap function are not 
necessarily at an angle $\pi/4$ from the crystallographic axes, and do not
in general form an angle of $\pi/2$. For the 
$BSCCO$ family the situation is 
different\cite{agl,plak}. The distortion occurs in such a way
that the new orthorhombic axes form an angle of $\pi/4$ with the original
tetragonal axes.
The relevant mixing is then between  $d_{xy}$ and $s$ or between 
$d_{x^2-y^2}$ and $s^-$ 
symmetries. The nodes are at the orthorhombic axes and, as in the pure 
$d$-wave case, the directions along which they occur form angles of 
$\pi/2$.
In the following sections,  we shall use these considerations to 
choose the gap functions to be studied.

In Section II we discuss methods to solve the nonlinear Maxwell-Equations 
in the Meissner regime to compute the physical quantities of interest. 
The general procedure we use exploits the fact, proved in Ref. \onlinecite{zv}
(see particularly Fig. 5 there) that the angular dependence of the
transverse magnetic moment for a typical finite sample (a flat crystal)
can be very accurately computed by assuming that the sample is infinite
in the $a-b$ plane, while the overall amplitude is underestimated somewhat
by  this assumption. Since, as we shall see, the nodal
structure is directly reflected in the angular dependences, we can proceed
for our purposes here by considering a simple ``slab'' geometry. This allows
us to do the entire work analytically, treating the small nonlinear
effects perturbatively.
We illustrate our perturbation method in Section II by considering first
a superconducting gap 
with mixed $d+s$ symmetry. 
Various additional gap functions, with nodes, are studied in Section 
III. We show how to include the effects of orthorhombic
distortion as they affect both
normal (FS shape) and superconducting (penetration depth tensor)
properties, in the calculation of the nonlinear
effects.  In Section IV we consider gaps without nodes in contrast with the
corresponding results obtained in the previous sections. We propose 
several ways that would allow to experimentally distinguish nodes from 
small minima in the superconducting gap. In the last Section we 
give our conclusions and possible guidelines for the future work.
\section{Methods}
\label{methods}
\subsection{Maxwell-London Electrodynamics}
We briefly review here the
nonlinear Maxwell-London equations\cite{sv,ys,zv,london} that
need to be solved 
for superconductors with  
various pairing states. For a $d$ wave infinite superconductor,
these nonlinear equations have been solved,
perturbatively\cite{ys} at zero temperature,
or numerically\cite{sv,zv} when finite systems or temperatures have been
considered.  

Here we are concerned with node spectroscopy, that is, with the effect
that a pattern of nodes (or deep dips) in the angular dependence of the
OP has on the dependence of the nonlinear Meissner effect
on the angle between applied field and sample orientation. For this
purpose we can work at $T=0$ since it was shown\cite{sv} that at the
temperatures $T<<T_c$ and field ranges 
 experiments are performed, the
$T=0$ result is sufficiently accurate. Similarly, we can work in a simple 
``infinite slab'' geometry (as defined below) since it is known\cite{zv}
that finite size effects, while changing the amplitude of the nonlinear
phenomena to some extent, do not affect their angular dependence. 
We will, therefore, develop a zero temperature perturbation
methods, for a simple geometry, that can be used with a wide variety of 
pairing states.
These methods are a nontrivial extension of those used
in Ref. \onlinecite{ys} for a pure $d$-wave state.

We consider a superconductor\cite{sv,ys} infinite in the $a-b$ plane 
and having a thickness $d$ in the $c$ direction. We will assume unless
otherwise stated that $d$ is larger than any relevant penetration
depth, (a ``slab'', not a film).  We 
use a coordinate system with $z$ axis parallel to the $c$ 
crystallographic direction. We follow the notation introduced in 
Ref.~\onlinecite{zv} unless otherwise indicated. 
The superfluid velocity ${\bf v}$ is defined as:
\begin{equation}
{\bf v}=\frac{\nabla \chi}{2} + \frac{e}{c} {\bf A},
\label{vdef}
\end{equation}
where $\chi$ is the phase of the superconducting OP, ${\bf 
A}$ the vector potential, and $e$ the proton charge, (with $\hbar=k_B=1$). 
The relation between ${\bf v}$ and ${\bf H}$ is given by  the
second London equation:
\begin{equation}
\nabla \times {\bf v}=\frac{e}{c}{\bf H}.
\label{londoneq}
\end{equation}
In the steady state the appropriate Maxwell equation is Amp\`ere's law,
$\nabla\times {\bf H}=\frac {4 \pi}{c}{\bf j}$.
Combining it with Eq. (\ref{londoneq}) we obtain:
\begin{equation}
\nabla\times\nabla\times{\bf v}=\frac{4 \pi e}{c^2}{\bf j(v)}.
\label{maxlon}
\end{equation}
The relation between ${\bf j}$ and ${\bf v}$ is generally               
 nonlinear and given by:\cite{ys,zv} 
\begin{equation}
{\bf j(v)}=-eN_f\int_{FS} d^2s \: n(s) {\bf v}_f [({\bf v}_f \cdot {\bf v})
  +2\int^{\infty}_0 d\xi \: f(E(\xi)+{\bf v}_f\cdot {\bf v})],
\label{nonlinjv}
\end{equation}
where  $N_f$ is the total density of states at the Fermi level,
$n(s)$ the density of states at point $s$ at the Fermi surface (FS),
normalized to unity ($\int_{FS} d^2s \: n(s)=1$), 
${\bf v}_f(s)$ the $s$-dependent Fermi
velocity, $f$ the Fermi function,
 with $E(\xi)=(\xi^2+\left| \Delta(s) \right|^2)^{1/2}$,
$T$ the absolute temperature and $\Delta(s)$ the OP.

For the geometry considered and for the  magnetic field 
${\bf H}_a$, applied in the $a-b$ plane, 
the fields have only $x$ and $y$ components, which depend   
 {\it only} on the coordinate $z$.  Eq. (\ref{maxlon}) then reduces to
\begin{equation}
\partial_{zz}{\bf v}+\frac{4 \pi e}{c^2}{\bf j(v)}=0, 
\label{lap}
\end{equation}                                                   
where it is understood that Eq. (\ref{nonlinjv}) is substituted in the
second term. The boundary conditions are:   
\begin{equation}
{\bf H}={\bf H}_a|_{z=\pm\frac{d}{2}}. 
\label{hatd}
\end{equation}
The first term in (\ref{nonlinjv}) is the usual relation
${\bf j}=-e \tilde{\rho} {\bf v}$, where $\tilde{\rho}$ is the superfluid
density tensor. At $T=0$
the nonlinear corrections (in ${\bf v}$),  described by the second
term of Eq. (\ref{nonlinjv}),  and due to the quasiparticle backflow can be 
written as:
\begin{eqnarray}
{\bf j}_{qp}{\bf (v)}&=& -2 eN_f\int_{FS} d^2s \: n(s) {\bf v}_f  
\Theta(-{\bf v}_{f} \cdot {\bf v}-\left| \Delta(s) \right| ) 
[({\bf v}_{f} \cdot {\bf v})^2-\left| \Delta(s) \right|^2]^{1/2} 
\nonumber \\
&\approx&-2e \sum_n N_{fn}\int_{\Omega_n} \frac{d \phi_n}{2 \pi}  
{\bf v}_{fn} 
[({\bf v}_{fn} \cdot {\bf v})^2-\left| \Delta(\phi_n) \right|^2]^{1/2}, 
\qquad n=1,2,...
\label{jq}
\end{eqnarray}
where we have replaced integration over the whole FS by integration 
over only  the nodal (or quasinodal) regions $\Omega_n$, each region
enumerated by the index $n$, where  the gap vanishes (or dips to a value
much smaller than its maximum) so that 
a backflow  current due to  so called\cite{ys} ``jets'' of quasiparticle 
excitations is allowed. 
Quantities at the node $n$ are denoted by an index ``$n$''. For 
an  anisotropic FS surface some care has to be taken: at the nodal 
position $s_n$, the density of states, $n(s_n)$, is not necessarily equal to 
the average value, $n_{iso}$. At each node we therefore define the    
appropriate weighted density of states $N_{fn} \equiv N_f n(s_n)/n_{iso}$.
 Regions contributing to the backflow     
current are determined by
\begin{equation}
\left| \Delta(\phi_n) \right|   +{\bf v}_f \cdot {\bf v} <0,
\label{reg}
\end{equation}
where $\phi_n$ is angle with respect to the closest node, $n$,  in the 
energy gap,
and  ${\bf v}_f$ can  
be approximated by ${\bf v}_{fn}$, its value at that node.
As an example, for a pure
$d$-wave ($d_{x^2-y^2}$) OP, 
$\Delta(\phi)=\Delta_d \sin(2\phi)$, where $\phi$ is 
measured with respect to a node,
Eq. (\ref{jq}) can be integrated with $\Omega_n$ 
given from Eq. (\ref{reg}) by $\left|\phi_n \right| \leq \phi_{c n}$,  
where $\phi_{cn}$ is determined from: 
\begin{equation}
({\bf v}_{fn} \cdot {\bf v})^2=\left| \Delta_d \sin 2 
\phi_{cn} \right|
\approx 4 \Delta_d^2 \phi_{cn}^2,  \label{phic}
\end{equation}
and  $\phi_{c n} \ll 1$,  
which implies that the nodal value is an accurate approximation on 
$\Omega_n$. It follows from these considerations that the nonlinear 
effects depend on the {\it local} values of FS quantities at the nodes. 
The overall shape of the FS enters (as we shall see in more detail 
below) very indirectly, through the eigenvalues of the London 
penetration depth tensor.  
\subsection{Nonlinear Currents}
To perform the calculation of the angular dependence of the nonlinear
electrodynamic quantities we must first develop a general method
to compute the current, including the nonlinear terms
from  Eq. (\ref{jq}), for the different
OP's under consideration. The resulting expressions lead,
after substitution into (\ref{lap}), to nonlinear differential equations that
can be solved perturbatively. 
It is
pedagogically convenient to illustrate both steps by carrying them out in
terms of 
a mixed $d+s$-wave gap function and an isotropic FS. It is then rather
straightforward to extend the results 
to the other
forms of the OP and to include the effects of the $a-b$ plane 
anisotropy.

The specific form of a $d+s$ OP that we consider in this paper is:
\begin{equation}
\Delta_{d+s}(\phi)=\Delta_s+\Delta_d \sin 2\phi, \label{dsgap}
\end{equation}
where all quantities are real, $\phi$ is measured from the $X$ axis 
(Fig. \ref{fig1}), and
we take $|\Delta_s| \ll |\Delta_d|$, which ensures that there are nodes 
in the gap function. The experimental constraints on the smallness of 
$\Delta_s/\Delta_d$ have been discussed, for example, by Annett et al.
\cite{agl,agl2}. At $\Delta_s=0$ we will recover the
known perturbative results\cite{ys} for that limit.
The relevance of this form of the gap to YBCO, when the effects of
orthorhombicity are taken into account, has been discussed above.
For
$\Delta_s \neq 0$, the nodes of Eq. (\ref{dsgap}) are no longer separated 
by an angle of $\pi/2$ since they are shifted by an angle $\pm\nu$ (see 
Fig. \ref{fig1}),
\begin{equation}
\nu\equiv \frac{1}{2} \sin^{-1} (\frac{\Delta_s}{\Delta_d}), 
\label{nu}
\end{equation}
from the orthogonal axes, which we denote by $X$ and $Y$,
and which will remain defined throughout the paper as being along the nodal
directions of a 
pure $d$-wave gap, that is, at angles of $\pi/4$ with the crystallographic
axes of YBCO.  These axes, together with  other details, are shown in 
Fig. \ref{fig1}. We 
 introduce also 
nonorthogonal axes along the nodal directions of the $d+s$ gap function with
unit vectors $\hat{x}_n$ ($n$=1,2,3,4 as indicated in the Figure):
\begin{equation}
\hat{x}_1=-\hat{x}_3=c \hat{X} -s \hat{Y}, \qquad 
\hat{x}_2=-\hat{x}_4=- s \hat{X} + c \hat{Y},  \label{unit}
\end{equation}
where  $c\equiv \cos \nu$ and  $s\equiv \sin \nu$.  
The angles $\phi_n$, defined below Eq. (\ref{reg}), measure angular 
shifts from the nodes of $\Delta_{d+s}$, for example, 
$\phi_1=\phi+\nu$. Close to the nodal region, the gap has the form: 
\begin{equation}
\left |\Delta_{d+s}(\phi_n) \right| \approx \left |\mu \Delta_{eff} \phi_n 
\right|, \qquad n=1,2,...
\label{dsnode}
\end{equation}
where the effective amplitude of the gap function                  
 $\Delta_{eff}=(\Delta_d^2-\Delta_s^2)^{1/2}$ is slightly reduced from 
the pure $d$-wave limit, and $\mu=2$ in this case. 
The allowed region for quasiparticle excitations from Eq. 
(\ref{reg}) is $\phi^2_{n} \leq \phi^2_{cn}$ where
\begin{equation}
\phi^2_{cn}=\frac{({\bf v}_{fn} \cdot {\bf v})^2}
{\mu ^2 \Delta_{eff}^2}. 
\label{phicds}
\end{equation}
We will investigate the angular dependence of  physical quantities, 
when the  field is applied along different directions with respect to 
nodes in the gap function. We define, as in Ref. \onlinecite{zv}, 
$\psi$ to be the angle of ${\bf H}_a$ with the $+\hat{Y}$ direction.
This angle is also shown in Fig.~\ref{fig1}.
 From Eqs. (\ref{jq}) and (\ref{reg}) we see that as $\psi$ changes, 
various nodes contribute to  ${\bf j}_{qp}$.
Without $a-b$ plane FS anisotropy the integration 
yields (as a special case from Appendix \ref{cure}):
\begin{equation}
j_{qpi}=\frac{1}{2}e N_f \frac{v_f^3}{\mu \Delta_{eff}}[e_{1i}(\psi) 
v_{X}^2+e_{2i}(\psi) v_{X} v_{Y}+e_{3i}(\psi) v_{Y}^2], \qquad 
i=X, Y, \label{jqds}
\end{equation}
where the coefficients $e_{1,2,3\:i}(\psi)$ are defined for 
various ranges of $\psi$:             
\begin{mathletters}
\label{es}
\begin{equation}
e_{1X}=e_{3Y}=c^3-s^3, \:
e_{2X}=2 e_{3X}=2 e_{1Y}=e_{2Y}=-2 cs(c-s),  \: \: \: \psi \in [\nu, 
\frac{\pi}{2}-\nu]
\end{equation}
\begin{equation}
e_{1X}=-e_{3Y}=-(c^3+s^3), \: 
e_{2X}=-2e_{3X}=2e_{1Y}=-e_{2Y}=2 cs(c+s), \: \: \: \psi \in  
[\frac{\pi}{2}-\nu, \pi+\nu].
\end{equation}
\end{mathletters}
We can simply verify that for a pure $d$ wave case
$(c\equiv 1, s\equiv 0)$ these expressions correctly reduce\cite{zv} to:
\begin{equation}
j_{qpX,Y}=e \rho_{ab}
\frac{v_{X,Y}\left| v_{X,Y} \right|}{2 v_c}, \label{jqd}
\end{equation}
where $\rho_{ab}=\frac{1}{2}N_f v^2_{f}$ is the superfluid density, 
related to the penetration depth by $\lambda^{-2}_{ab}=\frac{4 \pi 
e^2}{c^2} \rho_{ab}$, 
and the characteristic velocity in that limit is $v_c\equiv \frac{\Delta_d}{v_{f}}$.
In general, we define      
\begin{equation}
v_c\equiv\Delta_{eff}/v_f, 
\label{vc}
\end{equation}
and a dimensionless velocity 
\begin{equation}
u_i\equiv\frac{v_i}{\mu v_c}, \qquad i=X,Y.
\label{udef}
\end{equation}
We also define the dimensionless field:
\begin{equation}
 h=\frac{H_a}{H_0},
\label{hdef}
\end{equation}
where we introduce the characteristic magnetic field as:
\begin{equation}
H_0=\frac{c \Delta_{eff}}{e \lambda_{ab} v_f}.
\label{h0}
\end{equation}
In the case of pure $d$-wave gap, Eq. (\ref{h0}) reduces\cite{zv} to
$H_0=\frac{\phi_0}{\pi ^2 \lambda_{ab} \xi_{ab}}$,
where $\phi_0$ is the flux quantum and the in-plane coherence length
is defined as $\xi_{ab}=\frac{v_f}{\pi\Delta_d}$. 
Then, recalling  (\ref{jqds}), we can write  (\ref{lap}) in  
dimensionless form as: 
\begin{equation}
\partial_{ZZ}u_i-u_i
+[e_{1i}(\psi) u_{X}^2+e_{2i}(\psi) u_{X} u_{Y}+e_{3i}(\psi) 
u_{Y}^2]=0,
\qquad i=X,Y.
\label{ueq}
\end{equation}
Here we have introduced the dimensionless coordinate $Z\equiv z/\lambda_{ab}$,
where $\lambda_{ab}$ is the penetration depth at any of the four
equivalent nodes. 
In the pure $d$-wave limit the above equations again
reduce to the known\cite{ys} result.
In the geometry considered  
in this paper $j_{qp i}$
and $u_i$ have definite parity with respect to the $Z$ 
coordinate and it is sufficient to solve the boundary value problem 
for half of the slab ($Z \geq 0$). The boundary conditions at the surface 
$Z=Z_s \equiv \frac{d}{2 \lambda_{ab}}$, from Eq. (\ref{londoneq}) are:
\begin{equation}
\label{bcds}
\partial_Z u_{X}|_{Z=Z_s}=h\frac{\cos \psi}{\mu}, \qquad 
\partial_Z u_{Y}|_{Z=Z_s}=h\frac{\sin\psi}{\mu}, 
\end{equation}
and odd parity of ${\bf u}$ requires:
\begin{equation}
u_{X,Y} \equiv 0|_{Z=0}. \label{bco}
\end{equation}

\subsection{Perturbation method}
In the Meissner regime where $h, u \ll 1$, Eq. (\ref{ueq}), with the 
boundary conditions (\ref{bcds}) and (\ref{bco}), can be solved 
perturbatively. We write $u_i(Z)$ as a sum of linear and 
quadratic parts in the small\cite{small} parameter $h$, 
$u_i(Z)=L_i(Z)+N_i(Z)$. The linear part then satisfies:
\begin{equation}
\partial_{ZZ} L_i(Z)-L_i(Z)=0, \label{leq}
\end{equation}
with $L_i$ satisfying the boundary conditions  (\ref{bcds}), (\ref{bco}). 
The solution is:
\begin{equation}
L_i(Z)=a_i \sinh(Z), \qquad i=X,Y \label{linsol}
\end{equation}
and from Eq. (\ref{bcds}):
\begin{equation}
\label{adef}
a_{X}=h \frac{\cos\psi}{\mu \cosh(Z_s)}, \qquad
a_{Y}=h \frac{\sin\psi}{\mu \cosh(Z_s)}.
\end{equation} 
The nonlinear, perturbation part, in $h$, $N_i$, satisfies :
\begin{mathletters}
\label{neq}
\begin{equation}
\partial_{ZZ}N_i(Z)-N_i(Z)+\bar{a}_i^2 \sinh ^2 (Z)=0,
\end{equation}
\begin{equation}
\bar{a}^2_i=[e_{1i}(\psi) a_{X}^2+e_{2i}(\psi) a_{X} a_{Y}+e_{3i}(\psi)  
 a_{Y}^2], \qquad i=X,Y
\end{equation}
\end{mathletters}
at $Z=Z_s$ from Eq. (\ref{bcds}) the boundary conditions are:
\begin{equation}
\partial_Z N_i|_{Z=Z_s}=0. \label{nbc}
\end{equation}
The local magnetic field at $Z=Z_s$ is equal to 
the applied field. At $Z=0$, we have from Eq. (\ref{bco})
that $N_i(0)\equiv 0$.
The complete solution to Eq. (\ref{neq}) is a 
sum of the homogeneous solution, $N_{hi}$ and a particular solution, $N_{pi}$:
\begin{mathletters}
\label{a12sol}
\begin{equation}
N_i(Z)=A_{1i} \sinh(Z)+A_{2i} \cosh(Z)+P_i \cosh(2 Z)+R_i \equiv 
N_{hi}(Z)+N_{pi}(Z), 
\end{equation}
\begin{equation}
A_{1i}=\frac{2}{3} \bar{a}_i^2 (\sinh(Z_s)-\tanh(Z_s)),  \: \: 
A_{2i}=-\frac{1}{3} \bar{a}_i^2, \: \:  
P_i= \frac{1}{3} R_i=-\frac{\bar{a}_i^2}{6},
\qquad i=X,Y .
\end{equation}
\end{mathletters}
where the coefficients in $N_{hi}$, $A_{1i}$, $A_{2i}$, are obtained 
using Eq. (\ref{bco})   and (\ref{nbc}). 
The magnetic field can be calculated from the field 
${\bf u}$ through Eq. (\ref{londoneq}):
\begin{mathletters}
\label{hsol}
\begin{equation}
H_{X}(Z)= -H_0[h f_L(Z) \sin \psi + 
\frac{h^2}{2}
(e_{1Y}\cos^2 \psi+e_{2Y}\cos \psi \sin \psi+e_{3Y}\sin^2\psi)f_N(Z)], 
\end{equation}
\begin{equation}
H_{Y}(Z)= +H_0[h f_L(Z) \cos \psi +
\frac{h^2}{2}
(e_{1X}\cos^2 \psi+e_{2X}\cos \psi \sin \psi+e_{3X}\sin^2\psi)f_N(Z)]
\end{equation}
where
\begin{eqnarray}
f_L(Z)&=&\frac{\cosh(Z)}{\cosh(Z_s)}, \\
f_N(Z)&=&\frac{1}{3 \cosh^2(Z_s)}
[\:2(\sinh(Z_s)-\tanh(Z_s)) \cosh(Z)+2\sinh(Z)-\sinh(2 Z)]. \nonumber
\end{eqnarray}
\end{mathletters}
The above results are valid for any value of $d$ (not just  
$(d \gg \lambda)$),  
and we can at once write expressions for measurable quantities. The
components of the penetration depth along the two orthogonal directions
defined by the axes can be 
obtained as:
\begin{equation}
\label{pen}
\frac{1}{\lambda_{X}({\bf H}_a)}=\frac{1}{H_a}|\partial_z H_{X}|_{Z=Z_s}, 
\qquad
\frac{1}{\lambda_{Y}({\bf H}_a)}=\frac{1}{H_a}|\partial_z H_{Y}|_{Z=Z_s}.
\end{equation}
For a slab, $Z_s \gg 1$ $(d \gg \lambda)$, a pure $d$-wave gap function 
($\nu=0$) and field applied along the node or antinode, one can use these
and similar expressions to  recover the   
results for $1/\lambda({\bf H}_a)$ from  Refs.~\onlinecite{sv,ys}.

We next consider the magnetic moment, in particular its component 
transverse to the applied magnetic field, which in this case is
purely nonlinear for a symmetric sample\cite{demag} . 
In previous work\cite{zv,jcp} we have shown that the magnetic moment 
(for any sample geometry) can be calculated using only the 
 surface values of the fields. For the geometry 
considered here, the  magnetic moment, defined in general as:
\begin{equation}
{\bf m}= \frac{1}{2 c} \int 
d {\bf r} {\bf r} \times {\bf j(v)},
\label{mom}
\end{equation}
can be expressed as:
\begin{equation}
m_{x,y}=-\frac{{\cal S} d \:H_{a\:x,y}} {4 \pi} \mp 
\frac{{\cal S} c}{2 \pi e} v_{y,x}(d/2), \label{msur}
\end{equation}
where $x$ and $y$ are orthogonal axes fixed in space, ${\cal S}$ is the slab 
area, and
we have used that ${\bf v}$ is odd in $z$\cite{brandt}. 
For a slab, $(d\gg \lambda_{ab})$, which is the case of experimental
interest\cite{sv}, the     
expressions for $N_{X,Y}$ from Eq. (\ref{a12sol}) simplify:
\begin{equation}
N_i(Z_s)=\frac{h^2}{3 \mu^2}(e_{1i} \cos^2\psi+ e_{2i} \cos \psi \sin \psi
+e_{3i}\sin^2 \psi) \qquad i=X,Y
\label{nsur}
\end{equation}
 and we obtain the transverse magnetic moment, 
$m_\bot=m_{X} \cos \psi +m_{Y} \sin \psi$:
\begin{eqnarray}
\label{mbig}
m_\bot(\psi)&=&\frac{{\cal S} \lambda_{ab}}{3 \mu^2\pi}  \frac{H_a^2}{H_0}
[ e_{3X} \sin^3 \psi-e_{1Y}\cos^3 \psi \\
&+&\cos \psi \sin\psi ((e_{1X}- e_{2Y}) \cos \psi+
(e_{2X}-e_{3Y}) \sin \psi)]. \nonumber
\end{eqnarray}
The most obvious consequence of the $s$-wave admixture is that
the periodicity of
$m_\bot(\psi)$ is no longer strictly  $\pi/2$, but $\pi$, although the 
$\pi/2$ Fourier component remains larger for small $\nu$. 
Writing explicitly $e_{1,2,3\:i}$  we have: 
\begin{mathletters}
\label{mpsi}
\begin{equation}
m_\bot(\psi)=\frac{{\cal S} \lambda_{ab}}{12\pi} \frac{H_a^2}{H_0}
(c-s)(\cos \psi -\sin \psi) [c s+(1+4 cs) \cos\psi \sin\psi], \: \: \:
\psi \in [\nu, \frac{\pi}{2}-\nu]
\end{equation}
\begin{equation}
m_\bot(\psi)=-\frac{{\cal S} \lambda_{ab}}{12\pi} \frac{H_a^2}{H_0}
(c+s)(\cos \psi+\sin \psi) [c s+(1-4 cs) \cos\psi \sin\psi], \: \: \:
\psi \in [\frac{\pi}{2}-\nu, \pi+\nu].
\end{equation}
\end{mathletters}
We can readily extend this
solution for $m_\bot(\psi)$ to the remaining
$\psi$ range by symmetry. 
In the limit $\nu \rightarrow 0$, corresponding to 
a pure $d$-wave, we recover the known result for this limit\cite{yseqn}.

The above expressions are in terms of the angle $\psi$ with a line
of nodes. In an experimental situation the position of the line of
nodes is not {\it a priori} known, since it depends on the amount
of $\Delta_s$ admixture, which is to be determined. All one has to 
do is to plot $m_\perp$ as a function
of angle, with arbitrary origin, and compare the shape of the curve,
up to an horizontal translation, with the plots presented below.

We now show in Fig. \ref{mspf}
the angular dependence of $m_\bot(\psi)$ from Eq. 
(\ref{mpsi}) for  $\psi \in [-\frac{\pi}{4}, \frac{3\pi}{4}]$. 
This interval is a 
complete
period for the plotted quantity. The solid line represents 
the pure $d$-wave result, with its maximum normalized to unity. 
The actual unnormalized value of its amplitude as a function of 
applied field and sample parameters can be read off Eq. (\ref{mpsi}). 
It is discussed in detail, with inclusion of finite size effects, in 
Refs. \onlinecite{zv,jcp}, where typical values of $H_0$, $h$, ${\cal 
S}$, and other sample parameters, are given. The other  lines 
depict $m_\bot(\psi)$ for a $d+s$ gap. They are normalized 
by the same factor as the solid line, so that the 
relative effects of introducing an $s$ component can be gauged.
These curves, which are (see Eq. (\ref{mpsi})) independent of the applied 
field, are labeled by the value of the angle $\nu$, the shift of the 
nodal directions from the $d$-wave nodes. 
The relation between $\nu$ and $\Delta_s/\Delta_d$ is 
given by Eq. (\ref{nu}). If we take $\nu <0$ ($d-s$ gap) the lines are 
simply shifted by $\pi/2$. In Fig. \ref{mspf} we see that when the
nodal directions 
are not orthogonal, and hence the period of $m_\bot$ doubles from  its
$d$-wave value, the formerly equivalent values of the maxima and
minima of $\left|m_\bot(\psi)\right|$ take increasingly 
different values. One of the maxima becomes now the primary maximum, and its
value increases with $|\nu|$, while the value of the other maximum decreases 
with $|\nu|$. The overall increase in the maximum nonlinear effect occurs
largely because the contributions from the two quasiparticle ``jets'' 
at each node
tend to cancel each other to a lesser extent when the
angle between them is larger than $\pi/2$, that is, near the primary
maximum.  From the symmetry of the $d+s$ 
gap function, for any $\nu$ (and isotropic FS), when the field is 
applied along antinodes, ${\bf v} \: \| \: {\bf j}$ and             
 $m_\bot(\psi)\equiv0$ when $\psi=2(n+1)\pi/4$, $n=0,1,2,..$ (for 
$d$-wave this is also true for the field applied along the nodes).

For experimental analysis, it may be convenient to examine the
main Fourier components of the periodic $m_\bot(\psi)$ curve,
in particular its part
quadratic in the applied field $H_{a}$. 
As one can see from Fig. \ref{mspf}, the transverse moment has odd parity
with respect to $\psi$ if the angle is measured from a line of {\it antinodes}.
With that choice, only the coefficients of the sine terms in the Fourier 
analysis are nonvanishing. The ones corresponding to $\pi$ and $\pi/2$
periodicities are easily obtained 
from Eq. (\ref{mpsi}):
\begin{mathletters}
\label{fcomp}
\begin{equation}
m_2=\frac{{\cal S}\lambda_{ab}}{3\pi} \frac{H_a^2}{H_0}
[cs(1+\frac{1}{5} \cos 2\nu)+\frac{1}{12} \sin 2\nu-\frac{1}{20}\sin 
6\nu],
\end{equation}
\begin{equation}
m_4=\frac{{\cal S}\lambda_{ab}}{3\pi} \frac{H_a^2}{H_0}
[\frac{1}{4}-2c^2s^2-\frac{2}{15}\cos 4 \nu+\frac{1}{28} \cos 6 \nu+ 
\frac{1}{7} cs \sin 6\nu],
\end{equation}
\end{mathletters}
where $m_2$ and and $m_4$ are the $\pi$ and $\frac{\pi}{2}$
Fourier components of $m_\bot(\psi)$, respectively. 
When Fourier components are computed with respect to an
arbitrary angular origin, one will in general measure both sine and
cosine coefficients, but the square root of the sum of their squares
should equal the absolute value of the results given in Eq. (\ref{fcomp}).

In Fig. \ref{mspff} we
plot the Fourier components $m_2$, $m_4$,  as a function of 
$\nu$ and at a constant field, normalized so that $m_4$ is unity for 
$d$-wave. The solid line represents  $m_4$. We see that despite the
overall period having doubled, this Fourier component is only slightly reduced 
for the range of $\nu$ we consider\cite{agl2}. The
$\pi$ component, $m_2$, 
is depicted by the dotted line and it increases monotonically with $\nu$.

In the following sections we will focus on the nonlinear 
transverse magnetic moment for different forms of  gap 
functions. Other quantities could be obtained as shown above.
We will see that the generalization of the methods discussed in this Section
for the evaluation of the currents and for the perturbation values of the
fields is straightforward.

\section{Gap Functions With Nodes}
\label{node}
In this section we consider additional relevant order parameters with nodes.
We will include the details of the derivations in so far as
they differ from the example
discussed above, but the emphasis will be in the results for the angular
dependence of the transverse moment.

\subsection{Anisotropic d-wave states}
The first examples we consider are those of a pure $d$-wave order
parameter, with nodes at $\pi/2$ angles, but where the FS in the $a-b$ 
plane is anisotropic, so that the penetration depths 
are not equivalent. This is a possible state for both the YBCO and 
BSCCO families. For the latter case the situation is easier since the
Fermi velocities at the nodes are still aligned with the $X$ or $Y$
orthogonal axes. We therefore concentrate on the more complicated first
case, where the Fermi velocities at the nodes are not aligned with these
axes, but are equal in magnitude.
\subsubsection{YBCO type orthorhombicity}
For YBCO, although the orthorhombic distortion of the crystal 
lattice is small, there is a quite significant      
in-plane penetration depth anisotropy,\cite{ybco}
$(\lambda_a^2/\lambda_b^2  
\sim 2-6)$. Even for a pure $d$-wave 
gap function,  
the vectors ${\bf v}_{fn}$ no longer lie along the nodal 
directions $\pm \hat{X}, \: \pm \hat{Y}$,
but are shifted 
by an angle $\pm \nu_A$. We define $\nu_A$ as the angle between the 
$+\hat{Y}$ axis and  ${\bf v}_{f2}$. 
In Fig. \ref{axes} we show  relevant
quantities for an anisotropic FS and the more general case of a $d+s$ gap. 
The pure $d$-wave limit, considered in this
subsection, corresponds to setting $\nu\equiv0$ in the Figure.
We can write $\nu_A$ in terms of an anisotropy parameter $\Lambda_0$:
\begin{equation}
\nu_A\equiv  tan^{-1}(\frac{ \Lambda_0^2-1}{\Lambda_0^2+1}),
\label{nua}
\end{equation}
where $\Lambda_0$ can be evaluated in terms of Fermi surface parameters. For an
ellipsoidal FS one has
 $\Lambda_0=\frac{\lambda_a}{\lambda_b}$, 
but our results for the nonlinear currents are not restricted to this case, 
since they depend only on the properties at the nodes, not on the 
details of the  
shape of the FS. Then,
${\bf v}_{fn}= v_{fn} \hat{x}_n$ and $ v_{f1} \equiv 
v_{f2} \equiv v_{f3} \equiv v_{f4}$ with the directions given by:
\begin{equation}
\hat{x}_1=-\hat{x}_3=\frac{(c_A-s_A)}{\sqrt{2}}\hat{a} 
-\frac{(c_A+s_A)}{\sqrt{2}}\hat{b}, \: \: \:
\hat{x}_2=-\hat{x}_4= \frac{(c_A-s_A)}{\sqrt{2}}\hat{a} 
+ \frac{(c_A+s_A)}{\sqrt{2}}\hat{b},  \label{unitda}
\end{equation}
where $\hat{a}, \hat{b}$ are unit vectors along the $a$, $b$ 
crystallographic axes, $c_A\equiv\cos \nu_A$  and $s_A\equiv\sin \nu_A$.
The calculation is then quite analogous to that done in the previous
section. We can obtain, from Appendix 
\ref{cure}, $j_{qp 
\:i}$ as given in Eq. (\ref{jqds}) with $v_f$ being replaced by $v_{fn}$. 
The linear part is obtained most conveniently in terms of the
principal axes of the superfluid density tensor, which in this case
are the crystallographic directions. 
If we now define the dimensionless length $Z$
in terms of $\lambda_n$ $(\lambda_n^{-2}=\frac{2 \pi e^2}{c^2} N_{fn} 
v^2_{fn}$), the penetration depth at the nodes, (instead of    
$\lambda_{ab}$), $v_c\equiv\frac{\Delta_d}{v_{fn}}$ and $H_0\equiv\frac{c 
\Delta_d}{e \lambda_n v_{fn}}$, then by following
the steps leading to the derivation of (\ref{ueq}) and the discussion from 
Appendix \ref{cure}, we obtain:
\begin{equation}
\partial_{ZZ}u_i-\frac{\lambda_n^2}{\lambda_i^2}u_i
+[e_{1i}(\psi) u_{a}^2+e_{2i}(\psi) u_{a} u_{b}+e_{3i}(\psi) 
u_{b}^2]=0,
\qquad i=a,b
\label{ueqa}
\end{equation}
which is formally identical to (\ref{ueq}) and is solved in
the same way.
We can express the linear part of the velocity components along the
principal axes as:
\begin{mathletters}
\label{alinsol}
\begin{equation} 
L_i(Z)=a_i\sinh(r_i Z), \qquad i=a,b 
\end{equation}
\begin{equation} 
a_{a}=h \frac{\cos(\psi-\frac{\pi}{4})}{\mu r_a\cosh(r_a Z_s)}, \qquad
a_{b}=h \frac{\sin(\psi-\frac{\pi}{4})}{\mu r_b \cosh(r_b Z_s)},
\end{equation} 
\end{mathletters} 
where we have defined $r_i \equiv \lambda_n /\lambda_i$, $i=a,b$.
The nonlinear part satisfies: 
\begin{equation}
\partial_{ZZ}N_i-r_i^2 N_i
+[e_{1i} a_{a}^2 \sinh^2(r_a Z)+e_{2i} a_{a} a_{b} \sinh(r_aZ)\sinh(r_bZ)
+e_{3i} a_{b}^2 \sinh^2(r_bZ)], 
\label{neqa}
\end{equation}
with the solution $N_i(Z)\equiv N_{hi}(Z)+ N_{pi}(Z)$, where:
\begin{equation}
 N_{hi}(Z)=A_{1i}\sinh(r_i Z) + A_{2i} \cosh(r_i Z), \qquad i=a,b
\label{ahomo}
\end{equation}
\begin{equation}
N_{pi}(Z)=P_i \cosh(2 r_a Z)+ R_i \cosh((r_a+r_b)Z)
+S_i \cosh((r_a-r_b)Z)+ T_i \cosh(2 r_b Z)+U_i. 
\label{apart}
\end{equation}
After determining the unknown coefficients from the boundary  conditions, 
as explained in Appendix \ref{coef}, we obtain:
\begin{mathletters}
\label{anthick}
\begin{equation}
N_a(Z_s)=\frac{h^2}{12}
[e_{1a} \frac{\cos^2(\psi-\frac{\pi}{4})}{r_a^4}
+\frac{3e_{2a}\cos(\psi-\frac{\pi}{4})\sin(\psi-\frac{\pi}{4})}
{r_a^2 r_b(2r_a+r_b)}
+\frac{3e_{3a}\sin^2(\psi-\frac{\pi}{4})}{r_a r_b^2(r_a+2 r_b)}],
\end{equation}
\begin{equation}
N_b(Z_s)=\frac{h^2}{12}  
[\frac{3e_{1b}\cos^2(\psi-\frac{\pi}{4})}{r_a^2 r_b(2r_a+r_b)}
+\frac{3e_{2b}\cos(\psi-\frac{\pi}{4})\sin(\psi-\frac{\pi}{4}) }
{r_a r_b^2(r_a+2r_b)} 
+\frac{e_{3b}\sin^2(\psi-\frac{\pi}{4})}{r_b^4}].
\end{equation}
\end{mathletters}
In addition to the nonlinear transverse magnetic moment, quadratic in $H_a$, 
there is a transverse component
linear in $H_a$ due to the anisotropy of the   
FS. From Eq. (\ref{linsol}) and (\ref{msur})  we get for  the linear 
part, $m_\bot^{lin}(\psi)$: 
\begin{equation}
m_\bot^{lin}(\psi)=\frac{{\cal S}}{4\pi} H_a 
(\lambda_a-\lambda_b) \cos 2 \psi. 
\label{mlin}
\end{equation}
Experimentally, the different angular and, particularly, 
field dependence should provide a way to separate the effects of 
$m_\bot^{lin}$  from the quantity of interest, the nonlinear $m_\bot$ 
which reflects the symmetry of the superconducting gap, rather than that
of the normal state.
 From Eq. (\ref{msur}) and (\ref{anthick}) we can express the nonlinear part  
of $m_\bot$, as in the $d+s$ case, in terms of the angle $\psi$:   
\begin{mathletters}
\label{ampsi}
\begin{eqnarray}
m_\bot(\psi)&=&\frac{{\cal S} \lambda_n}{12\pi} \frac{H_a^2}{H_0}
(c_A-s_A)(\cos \psi -\sin \psi)
\frac{1}{4}[3 \frac{(c_A+s_A)^2}{r_a r_b^2(r_a+2r_b)}
-\frac{(c_A-s_A)^2}{r_a^4} \nonumber \\ 
 &+&(18\frac{(c_A+s_A)^2}{r_a r_b^2(r_a+2r_b)}
-2\frac{(c_A-s_A)^2}{r_a^4}) \cos\psi \sin\psi], \qquad 
\psi \in [\psi_1,-\psi_1+\frac{\pi}{4}]
\end{eqnarray}
\begin{eqnarray}
m_\bot(\psi)&=&-\frac{{\cal S} \lambda_n}{12\pi} \frac{H_a^2}{H_0}
(c_A+s_A)(\cos \psi+\sin \psi)  
\frac{1}{4}[3\frac{(c_A-s_A)^2}{r_a^2 r_b(2r_a+r_b)}-
\frac{(c_A+s_A)^2}{r_b^4}  \nonumber \\ 
&-&(18\frac{(c_A-s_A)^2}{r_a^2 r_b(2 r_a+r_b)}-2\frac{(c_A+s_A)^2}{r_b^4}) 
\cos\psi \sin\psi], 
\qquad \psi \in [-\psi_1+\frac{\pi}{4}, \psi_1+\pi].
\end{eqnarray}
\end{mathletters}
where 
$\tan(\psi_1-\frac{\pi}{4})
=\frac{\lambda_b}{\lambda_a}\tan(\frac{\pi}{4}+\nu_A)$, 
as explained in Appendix B.
The effect of FS anisotropy on $m_\bot(\psi)$ is similar to that due to
the 
presence of an $s$-wave gap, discussed in the previous section. The 
period is doubled from the pure $d$-wave and isotropic FS value.
The results from Eq. (\ref{ampsi}) can be simply extended to   
include an admixture of $s$-wave gap and we will defer their discussion 
for the following subsection.
\subsubsection{BSCCO type orthorhombicity}
Here we consider the nonlinear current response for a $d$-wave OP when there 
is a FS anisotropy consistent with the orthorhombicity
of materials in the BSCCO family. The 
effects of the orthorhombic distortion relevant to the OP symmetry are 
discussed in Refs. \onlinecite{agl,agl2}, and in the 
Introduction. The nodal lines of the $d_{x^2-y^2}$ 
gap functions are along the $X$, $Y$ axes.  
These  axes  are  now the principal orthorhombic directions.
 As a consequence, the Fermi speeds at the nodes are nonidentical
($v_{fX} \neq v_{fY}$) 
and the nonlinear current which is generically $\propto v_{fn}^3$  
will differ along the $X$, $Y$ directions. Hence, some small changes are
needed in the procedure used 
for the YBCO case. We define the unit of length in terms of 
$\lambda\equiv\sqrt{\lambda_{nX} \lambda_{nY}}$, where $\lambda_{ni}$, 
$i=X, Y$ is defined analogous to $\lambda_n$, given above Eq. 
(\ref{ueqa}).
We introduce $\lambda$ and $v_f\equiv\sqrt{v_{fX} v_{fY}}$,
 in the previous definitions of 
the dimensionless units, for example, in  $Z\equiv z/\lambda$ and in $H_0$ 
from Eq. (\ref{h0}). We define 
$q_i \equiv (\lambda/\lambda_i)^{1/2}$, $i=X,Y$ and obtain:
\begin{eqnarray}
\partial_{ZZ}u_i-&q&_i^2 u_i
+p^2_i[e_{1i}(\psi) u_{X}^2+e_{2i}(\psi) u_{X} u_{Y}
+e_{3i}(\psi) u_{Y}^2]=0, \nonumber \\
&p&^2_i\equiv q^2_i \frac{\lambda_i^2}{\lambda_{ni}^2}\frac{v_{fi}}{v_f},
\qquad i=X,Y
\label{ueqbs}
\end{eqnarray}
where the only nonvanishing
coefficients $e_{1,2,3,i}$ are (cf. Eq. (\ref{jqd})),  
$e_{1X}=e_{3Y}=1$ for $\psi \in [0, \pi/2]$ and $-e_{1X}=e_{3Y}=1$ 
for $\psi \in [\pi/2, \pi]$ . 
The part of the velocity linear in $h$, $L_i$, has the form given by Eq. 
(\ref{alinsol})  with $r_i$  replaced by $q_i$ and $\psi-\frac{\pi}{4} $ 
 by $\psi$. 
The nonlinear part in $h$, $N_i$,  from Eq. (\ref{ueqbs}) satisfies :
\begin{mathletters}
\label{neqbs}
\begin{equation}  
\partial_{ZZ}N_i(Z)-q_i^2 N_i(Z)+\bar{a}_i^2 \sinh ^2 (q_iZ)=0,
 \qquad i=X,Y
\end{equation}
\begin{equation}
\bar{a}^2_X=p_{X}^2 e_{1X} a_{X}^2, \qquad  
\bar{a}^2_Y=p_{Y}^2 e_{3Y} a_{Y}^2, 
\end{equation}
\end{mathletters}
It is straightforward, following the $YBCO$ case, to write the complete 
solution for $N_i$ and obtain for a slab:
\begin{equation}
N_X(Z_s)=\frac{h^2}{12}\frac{p_{X}^2}{q_{X}^2}e_{1X} \cos^2\psi, \qquad
N_Y(Z_s)=\frac{h^2}{12}\frac{p_{Y}^2}{q_{Y}^2}e_{3Y} \sin^2\psi.
\label{nbs}
\end{equation}
The corresponding nonlinear transverse magnetic moment is
\begin{equation}
m_\bot(\psi)=\frac{{\cal S} \lambda}{12 \pi} 
\frac{H^2_a}{H_0} \cos \psi \sin \psi 
[e_{1X}\frac{p_{X}^2}{q_{X}^2} \cos \psi -e_{3Y} 
\frac{p_{Y}^2}{q_{Y}^2}\sin \psi],
\label{mbs}
\end{equation}
where $e_{1X}$ and $e_{3Y}$ are as given below Eq. (\ref{neqbs}).
We see that the two terms, which correspond to  two ``jets'' of 
quasiparticle backflow excitations in the nodal directions are of 
different strengths since $p_{X}/q_{X} \neq 
p_{Y}/q_{Y}$. We illustrate this effect 
on the angular dependence of $m_\bot$ in Fig. \ref{mbscof}. 
We use here the same conventions as in Fig. \ref{mspf}.
The
solid line  represents results obtained for $m_\bot(\psi)$ in the isotropic 
FS case,  with its maximum  normalized 
to unity. They should be contrasted with the broken line  
 which corresponds to results with  
$\lambda_{Y}/\lambda_{X}=1.1$,\cite{timusk}
normalized by the same factor.
 We have used 
Ref. \onlinecite{oda} to estimate $\lambda_i/\lambda_{ni} \sim \sqrt{2}$, 
and set $v_{fX}/v_{fY}=\lambda_{Y}/\lambda_{X}=1.1$. Half a period 
for $m_\bot$ is shown, the plot can be extended by odd parity over the whole 
range, $\psi \in [-\pi/2, \pi/2]$. The behavior of $m_\bot$ is somewhat 
similar to that for a $d+s$ gap and isotropic FS (Fig. \ref{mspf}), but instead 
of $m_\bot$ vanishing for a field applied along the antinodes it vanishes 
here when the field is applied along the nodal directions.

\subsection{Anisotropic d+s}
The derivation of (\ref{ampsi}) can be combined with that of
(\ref{mpsi}) to include an  $s$-wave 
component in the energy gap, i.e. $\Delta_{d+s}(\phi)$ from Eq. 
(\ref{dsgap}), in the presence of FS anisotropy
of the YBCO type.  Examination of the derivations
of the equations just mentioned shows that the crucial point
about the additional $s$ term is that the shifts in the direction of
the ${\bf v}_{fn}$ 
from the $\pm \hat{X}$, $\pm \hat{Y}$ axes, arising from the anisotropy 
and from  the $s$ component
must be combined. Thus the total shift  angle $\nu_A$, introduced in
Fig. \ref{axes}, must  include the effect of $\nu$ as 
well as that of $\Lambda_0^2$. For an ellipsoidal FS this is achieved by 
replacing $\Lambda_0^2$ in 
Eq.~(\ref{nua}) with:
\begin{equation}
\Lambda^2 =\Lambda_0^2 \tan (\frac{\pi}{4}+\nu), 
\label{nuas}
\end{equation}
and we note that if $\nu$ is small and $\Lambda_0^2 \approx 1$ then 
$\nu_A \approx \nu_A^0+\nu$, where $\nu_A^0$ is the contribution 
arising solely from FS anisotropy. 
With these generalizations, 
the results for $m_\bot(\psi)$ in terms of $\nu_A$ are still given by 
Eq. (\ref{ampsi}).
Some care has to be taken 
with redefining dimensionless quantities. 
The characteristic 
velocity is taken to be $v_c=(\Delta_d^2-\Delta_s^2)^{1/2}/v_{fn}$.

In Fig. \ref{mybcof} we show the results from  Eq. 
(\ref{ampsi}) for the nonlinear transverse magnetic moment
$m_\bot(\psi)$ (the linear part is given by Eq. (\ref{mlin})). 
In the Figure, we give results for  $\Lambda_0\equiv\lambda_a/\lambda_b=1.5$ 
(which is in the correct range\cite{ybco} for YBCO) and take 
$\lambda_n^2=\lambda_a \lambda_b$. 
Again, the solid line is our standard reference result: pure $d$-wave 
and isotropic FS, normalized to unity. The broken lines are labeled by 
their $\nu_A$ values and are normalized by the same factor.
For the value of $\Lambda_0$ 
employed, the range of $\nu_A$ that we have chosen to display would 
correspond, for an ellipsoidal FS, to slightly negative values of 
$\nu$. This seems physically appropriate to describe YBCO, where the 
chains are in the $b$ direction and $\lambda_a > \lambda_b$. The 
departure of $m_\bot(\psi)$ from $\pi/2$ periodicity can be very 
significant when the effects of $\nu$ and $\nu_A^0$ are in the same 
direction.
The precise shape of $m_\bot(\psi)$ is the result of the combined
effects of FS 
anisotropy, reflected in the penetration depth, and of the $s$ wave
contribution. Experimental results for $m_\bot(\psi)$ would yield 
$\nu_A$, and information about the FS (i.e., $\Lambda_0$) would be needed 
to infer $\nu$. 

\subsection{Extended s-wave states}
The results we have discussed so far were obtained for nodes along four
directions on the FS. It is
straightforward to include 
symmetries of the OP with a
larger number of nodes, as we have already mentioned in the derivation
leading to (\ref{jq}). Among the allowed spin singlet 
pairing states\cite{agl} of a single $CuO_2$ plane with square lattice 
structure that have not been considered yet, are the
$A_{2g}$ or $``g$-wave'', and the ``extended $s$-wave'' states, 
with eight nodes.\cite{eight} The gap 
function for the former case can be written as 
\begin{equation}
\Delta_{es}(\phi)= \Delta_s + \Delta_g \sin 4\phi,
\label{esgap}
\end{equation}
where $|\Delta_s|\ll|\Delta_g|$. The $g$-wave is the limit $\Delta_s\equiv 0$.
We have used an angle $\phi$ measured from the $g$-wave nodal 
direction, along the $+X$ axis
(more conventionally,  
for an angle measured from 
the $a$-axis, the sine in Eq. (\ref{esgap}) would have been replaced by 
a cosine, and reflect the same symmetry properties\cite{agl} of 
$\Delta_{es}$ and the simple $s$-wave gap for a square lattice).
The   differences between  pure 
$g$-wave and extended $s$-wave, written in this form, are similar 
to those between $d$ and $d+s$ gap functions as explored in Section 
\ref{methods}. For example, the nodes are shifted from the $g$-wave gap by 
an angle $\pm \nu$, where we have:
\begin{equation}
\nu\equiv \frac{1}{4} \sin^{-1} (\frac{\Delta_s}{\Delta_g}), 
\label{nue}
\end{equation}
and $\nu$ is the angle between the direction of  
closest node to the $+X$ axis and the $+X$ axis.
The gap function at the nodes can be approximated by
\begin{equation}
\left| \Delta_{es}(\phi_n) \right| \approx 
\left| 4(\Delta_g^2-\Delta_s^2)^{1/2} 
\phi_n\right|,
\label{esnode}
\end{equation}
with $\phi_n$ taken with the respect to the node $n$ of $\Delta_{es}(\phi)$.
If we take an isotropic FS,
the Fermi velocity at the nodal directions is given by $v_{fn} \hat{x}_n$
with
\begin{eqnarray}
&\hat{x}_1&=-\hat{x}_5=c\hat{X}-s \hat{Y}, \: \: 
\hat{x}_2=-\hat{x}_6=c_{\alpha} \hat{X}+s_{\alpha} \hat{Y},\nonumber \\
 &\hat{x}_3&=-\hat{x}_7=s \hat{X}+c \hat{Y}, \: \:
\hat{x}_4=-\hat{x}_8=-s_{\alpha} \hat{X}+c_{\alpha} \hat{Y},
\label{esunit}
\end{eqnarray}
where $c \equiv \cos \nu$,  $s \equiv \sin \nu$,
$c_{\alpha} \equiv \cos \alpha$,  $s_{\alpha} \equiv \sin \alpha$,  and 
$\alpha \equiv \pi/4+\nu$.

After obtaining the nonlinear current by performing the sum in
(\ref{jq}) over all eight nodes,
the solution for the velocity field is readily found as in the $d+s$ case, 
noting that now $\mu \equiv 4$ and the coefficients $e_{1,2,3 \: i}$ are 
different.
The period of $m_\bot(\psi)$ 
becomes $\pi/2$.
The nonlinear current is of the form (\ref{jqds}) with 
\begin{mathletters}
\label{ees}
\begin{eqnarray}
e_{1X}&=&c^3-s^3+c^3_{\alpha}+s^3_{\alpha}, \: \: 
e_{2X}=2(-c s^2-c^2 s +c^2_{\alpha} s_{\alpha} 
- c_{\alpha} s^2_{\alpha})=2 e_{1Y}, \\  
e_{3X}&=&c s^2-c^2 s +c^2 _{\alpha} s_{\alpha}
+c_{\alpha}s^2_{\alpha}= \frac{1}{2}e_{2Y}, \: \:
e_{3Y}=-c^3-s^3-c^3_{\alpha} +s^3_{\alpha}, 
\qquad \psi \in [\nu, \frac{\pi}{4}-\nu] \nonumber 
\end{eqnarray}
\begin{eqnarray}
e_{1X}&=&c^3+s^3+c^3_{\alpha} +s^3_{\alpha}, 
 \: \:
e_{2X}=2(c s^2-c^2 s +c^2_{\alpha} s_{\alpha}
-c_{\alpha} s^2_{\alpha})=2 e_{1Y},  \\
e_{3X}&=&c s^2+c^2 s +c^2_{\alpha} s_{\alpha}
+c_{\alpha} s^2_{\alpha}= \frac{1}{2}e_{2Y}, \: \:
e_{3Y}=c^3-s^3-c^3_{\alpha}+s^3_{\alpha}, 
\qquad \psi \in [\frac{\pi}{4}-\nu,\frac{\pi}{2}+\nu]. \nonumber 
\end{eqnarray}
\end{mathletters} 
With the aid of Eq. (\ref{mbig}) we can obtain the corresponding 
$m_\bot(\psi)$:
\begin{mathletters}
\label{msg}
\begin{eqnarray}
m_\bot(\psi)&=&\frac{{\cal S} \lambda_{ab}}{48\pi} \frac{H_a^2}{H_0}
[(c+s)(cs-\frac{s}{\sqrt{2}}(c-s)) \cos^3 \psi
+(s-c)(cs-\frac{c}{\sqrt{2}}(c+s)) \sin^3 \psi  \\ \nonumber 
&+& \cos \psi \sin \psi 
[((c-s)(1+3cs)+\sqrt{2}c(1-\frac{3}{2}(c^2-s^2)))\cos \psi \\ 
&+&((c+s)(1-3cs)-\sqrt{2}s(1+\frac{3}{2}(c^2-s^2)))\sin \psi]], 
\qquad \psi \in [\nu, \frac{\pi}{4}-\nu]
\nonumber 
\end{eqnarray}
\begin{eqnarray}
m_\bot(\psi)&=&\frac{{\cal S} \lambda_{ab}}{48\pi} \frac{H_a^2}{H_0}
[(c-s)(cs+\frac{s}{\sqrt{2}}(c+s)) \cos^3 \psi
+(c+s)(cs+\frac{c}{\sqrt{2}}(c-s)) \sin^3 \psi \\ \nonumber 
&+& \cos \psi \sin \psi 
[((c+s)(1-3cs)+\sqrt{2}c(1-\frac{3}{2}(c^2-s^2)))\cos \psi \\  
&-&((c-s)(1+3cs)+\sqrt{2}s(1+\frac{3}{2}(c^2-s^2)))\sin \psi]],
\qquad \psi \in [\frac{\pi}{4}-\nu,\frac{\pi}{2}+\nu]
\nonumber
\end{eqnarray}
\end{mathletters}
where in the first definition of $H_0$, in Eq. (\ref{h0}), we have replaced 
$\Delta_{eff}$ by $(\Delta_g^2-\Delta_s^2)^{1/2}$.
\subsection{Mixed d+g Gap Functions}
Mixed pairing states compatible with the presence of orthorhombic 
distortion also include\cite{agl,agl2} $d+g$-wave states, i.e.,
 ``$s^{-}+d_{xy}$'' 
for the $YBCO$ family and  ``$s^{-}+d_{x^2-y^2}$'' for $BSCCO$. 
We consider here only the first case with an
isotropic FS. The other cases can be treated similarly.
We again assume that the $d$-wave character is dominant 
and we denote the gap function we use, with $\Delta_d \gg \Delta_g$ :
\begin{equation}
\Delta_{d+g}(\varphi)=\Delta_d \sin 2 \varphi+ \Delta_g \sin 4 \varphi,
\label{dggap}
\end{equation}
where $\varphi$ is measured from the $a$ axis.
At the four nodes we can approximate the absolute value of
the gap function, 
needed to calculate $j_{qp}$, with :
\begin{equation}
\left|\Delta_{d+g}(\phi_n)\right|\approx  \left|2 (\Delta_d 
\pm 2 \Delta_g) \phi_n \right|.
\label{dgnode}
\end{equation}
We have again contributions from two 
inequivalent quasiparticle ``jets'', in this case due to the different values  
of the gap near the nodes. The solution for the linear 
part of the velocity is the same as for the pure $d$-wave, discussed in the 
previous Section. If we define $u_i\equiv v_i/2 v_c$, $v_c\equiv 
\Delta_d/v_f$ the nonlinear part of velocity satisfies:
\begin{mathletters}
\label{neqdg}
\begin{equation}
\partial_{ZZ}N_i(Z)-N_i(Z)+\bar{a}_i^2 \sinh ^2 (Z)=0,
 \qquad i=X,Y
\end{equation}
\begin{equation}
\bar{a}^2_X=\frac{1}{1+2 r} e_{1X} a_{X}^2, \qquad  
\bar{a}^2_Y=\frac{1}{1-2 r} e_{3Y} a_{Y}^2, 
\end{equation}
\end{mathletters}
where $r \equiv \Delta_s/\Delta_d$, and $a_X$, $a_Y$ are given by Eq. 
(\ref{adef}).
The solution of Eq. (\ref{neqdg}) for a  slab is readily obtained 
as:
\begin{equation}
N_X(Z_s)=\frac{h^2}{12} \frac{e_{1X}}{(1-2 r)} \cos^2\psi, \qquad
N_Y(Z_s)=\frac{h^2}{12} \frac{e_{3Y}}{(1+2 r)} \sin^2\psi.
\label{ndg}
\end{equation}
The corresponding transverse magnetic moment is
\begin{equation}
m_\bot(\psi)=\frac{{\cal S} \lambda}{12 \pi} \frac{H^2_a}{H_0}
 \cos \psi \sin \psi 
[\frac{e_{1X}}{1+2 r} \cos \psi -\frac{e_{3Y}}{1-2 r} \sin \psi],
\label{mdg}
\end{equation}
where $e_{1X}, e_{3Y}$ are given from Eq. (\ref{jqd}). This result is 
formally similar to that for an anisotropic $d$-wave for BSCCO, 
recall Eq. (\ref{mbs}) and Fig. \ref{mbscof}, with the substitutions 
$p_{X}^2/q_{X}^2=1/(1+2 r)$ and $p_{Y}^2/q_{Y}^2=1/(1-2 r)$.
For brevity, we have not included plots in this and the previous 
subsections, but the interested reader can easily generate them from 
the analytic forms (\ref{msg}) and (\ref{mdg}).
\section{Nodeless Gap Functions}
\label{noless}
In this section we discuss the nonlinear supercurrent response when there 
are strictly speaking
no nodes but there are regions (``quasinodes'') with a very small although
finite superconducting gap.
For an example of a nodeless gap function of this
type  we choose $s+id$\cite{kot,joy} pairing. 
This is a frequently considered candidate for a 
gap function without nodes, and it clearly illustrates the difference in the 
nonlinear response when the nodes are absent and only  small minima    
in the gap function  exist. 
The range of $\left| \Delta_s/\Delta_d \right|$ that we 
consider is restricted to the very small values compatible
with the bounds set by  experiment\cite{agl2}. 
For these values, the quasiparticle contribution is dominated by
the behavior at the ``quasinodes'' and the
general procedures presented in the
previous sections can be used.
We will again concentrate here on 
the nonlinear magnetic moment, as a  probe to identify the 
symmetry of the gap function. The behavior of $m_\bot$ 
for the $s+id$ state is similar to that which would occur
for some other gap functions, such as an 
``anisotropic $s$-wave'', proposed in the 
framework of interlayer tunneling by Chakravarty et al.,\cite{chak} or 
for $d_{x^2-y^2}+i d_{xy}$\cite{lau}.

For brevity, we will assume an isotropic FS (equivalent nodes). 
Anisotropy can be added using the methodology of the previous section.
 We consider an $s+id$ gap function of the form
 $\Delta_{s+id}(\phi)=\Delta_s + i\Delta_d\sin(2 \phi)$, 
with $\Delta_s \ll \Delta_d$, where $\phi$ 
is the angle with respect to the $+X$ axis,  which is still 
minimum of $\left| \Delta_{s+id}(\phi) \right|$.
 The phase space available for 
quasiparticle excitations is reduced compared to the $d$-wave case 
so that the critical angle
(recall Eq. (\ref{reg})) is:
\begin{equation}
\phi^2_{cn}=\frac{({\bf v}_{f} \cdot {\bf v})^2-\Delta_s^2}
{4\Delta_d^2},  
\label{phicsid}
\end{equation}
There are no nonlinear effects present if $v < v_T\equiv \Delta_s/v_f$ 
or correspondingly if $H_a$ is smaller than a 
threshold field $h_T$ which we shall see is $h_T\equiv\Delta_s/\Delta_d$.
We can again perform an approximate integration  
of Eq. (\ref{nonlinjv}) and obtain an equation for the    
velocity field ($\psi \in [0, \frac{\pi}{2}]$):
\begin{equation}
\partial_{ZZ}u_i-u_i+(u_i^2-u_T^2) \Theta(u_i-u_T)=0, \qquad i=X, Y
\label{ueqsid}
\end{equation}
where $u_i=\frac{v_i}{2 v_c}$, $v_c$ is as defined for the $d$ 
wave state, and $u_T\equiv\frac{\Delta_s}{2 \Delta_d}$. The step function
$\Theta$ arises from  the nonlinear effects being present only for 
sufficiently large fields.

We proceed with the perturbation calculation as previously developed,
noting that we have to treat carefully  
the nonanalyticity  associated with the step
function\cite{pert}
in Eq. (\ref{ueqsid}). To do so, we divide the $Z \ge 0$ portion of the slab
(the solution for negative $Z$ follows from the symmetry) into  
two regions, the first
 where $u_i\ge u_T$ or correspondingly $Z\in [Z_{Ti}, Z_S]$  
and the second for $u_i < u_T$,  that is, for  $Z \in [0, Z_{Ti})$,
where the position $Z_{Ti}$ is to be determined. The solution for
the linear part $L_i(Z)$ is the same as in the $d$-wave case.
For the nonlinear 
part $N_i(Z)$ we have from (\ref{ueqsid}):
\begin{mathletters}
\label{nsid}
\begin{equation}
\partial_{ZZ}N_i-N_i+L_i^2-u_T^2=0, \qquad Z \in [Z_{Ti}, Z_S] \qquad 
i=X, Y
\end{equation}
\begin{equation}
\partial_{ZZ}N_i-N_i=0, \qquad Z \in [0, Z_{Ti}) \qquad i=X, Y.
\end{equation}
\end{mathletters}
$N_i(Z\leq Z_T)$ can then be expressed in the form of Eq. 
(\ref{leq}) and for $Z\leq Z_T$ we can write:
\begin{equation}
 N_i(Z)=A_{3i} \sinh(Z), \qquad i=X,Y
\label{ndown}
\end{equation}
where we have used $N_i(0)=0$. After finding the coefficients $P_i$ and
$R_i$ in the particular  
solution for $Z\geq Z_T$,  we are left with four unknown 
parameters $A_{1i}, A_{2i}, A_{3i}$ and the matching point $Z_{Ti}$. 
They are determined from Eq. (\ref{nbc}) and by requiring continuity 
of velocity, magnetic field, and current at $Z \equiv Z_{Ti}$. 
One has the same number of conditions and of unknowns.
Straightforward algebra yields:
\begin{mathletters}
\label{sid}
\begin{eqnarray}
A_{1i}=&-&\frac{a_i^2}{6} [\tanh(Z_s) \cosh(Z_{Ti}) (1-2 \sinh^2(Z_{Ti}))
 -4 \sinh(Z_s)] \\ \nonumber
&-&(\frac{a_i^2}{2}+u_T^2) \tanh(Z_S) \cosh(Z_{Ti}),
\end{eqnarray}
\begin{equation}
A_{2i}=\frac{a_i^2}{6} \cosh(Z_{Ti}) (1-2 \sinh^2(Z_{Ti}))
 +(\frac{a_i^2}{2}+u_T^2)\cosh(Z_{Ti}),
\end{equation}
\begin{eqnarray}
A_{3i}=&-&\frac{a_i^2}{6} [\tanh(Z_s) \cosh(Z_{Ti}) (1-2 \sinh^2(Z_{Ti}))
 -4 \sinh(Z_s)\\
&+&\sinh(Z_{Ti})(1+2 \cosh^2(Z_{Ti}))]    
  +(\frac{a_i^2}{2}+u_T^2)[\sinh(Z_{Ti})-\tanh(Z_S) \cosh(Z_{Ti})],\nonumber
\end{eqnarray}
\begin{equation}
P_i=-\frac{a_i^2}{6}, \qquad R_i=-\frac{a_i^2}{2}-u_T^2, \qquad 
 \sinh(Z_{Ti}(\psi))=\frac{u_T}{a_i(\psi)}, \qquad i=X,Y
\end{equation}
\end{mathletters}
where in the last expression we have emphasized the angular dependence 
of the matching point $Z_{Ti}$.  From
these equations we obtain that $N_i(Z_s=Z_{Ti})=0$. 
For the slab,  when         
$Z_s \gg 1$, the  expression for $N_i(Z_s)$ is simply:
\begin{mathletters}
\label{nsursid}
\begin{equation}
N_x(Z_s)=[\frac{1}{12} h^2 \cos^2 \psi -u_T^2+\frac{4 u_T^3}{3 h \cos \psi}]
\Theta(h-\frac{\Delta_s}{\cos \psi \Delta_d}),
\end{equation}
\begin{equation}
N_y(Z_s)=[\frac{1}{12} h^2 \sin^2 \psi -u_T^2+\frac{4 u_T^3}{3 h \sin \psi}]
\Theta(h-\frac{\Delta_s}{\sin \psi \Delta_d}).
\end{equation}
\end{mathletters}
We see from these expressions that for very small fields 
$h<h_T$, with $h_T\equiv \Delta_s/\Delta_d$ the nonlinear effects vanish.
For each jet of quasiparticles to be excited the applied field has to
be $h \geq h_{Ti}(\psi)$,  where $h_{Ti}(\psi)$ are
angle dependent threshold fields, the minimum value of which is the
overall threshold field $h_T$. If we now compute ${\bf m}_\bot$ as in
the derivation of Eq. (\ref{mbig}), we find:
\begin{eqnarray}
\label{msid}
m_\bot(\psi)&=&\frac{{\cal S} \lambda_{ab}}{12\pi} 
\:[\frac{H_a^2}{H_0} \cos\psi \sin \psi[\cos \psi \Theta(h-h_{TX})
-\sin \psi \Theta(h-h_{TY})]  \nonumber \\ 
&+&3(\frac{\Delta_s}{\Delta_d})^2 H_0
[\cos \psi \Theta(h-h_{TY})-\sin \psi \Theta(h-h_{TX})] \\ 
&-&2(\frac{\Delta_s}{\Delta_d})^3\frac{H_0^2}{H_a}
[\cot \psi \Theta(h-h_{TY}) -\tan \psi \Theta(h-h_{TX})]], \nonumber
\end{eqnarray}
where $h_{TX}$ and $h_{TY}$ are the threshold fields along the $X$ and 
$Y$ directions respectively, which can be read off Eq. (\ref{nsursid}).
It is apparent from this result that in this case $m_\bot$ has a
more  complicated angular and field dependence than in the $d$-wave 
gap (the limit $u_T\equiv \frac{\Delta_s}{2 \Delta_d} \rightarrow 0$ 
corresponds to the pure $d$-wave), or indeed than in any of the other
cases considered in the previous sections. 
In contrast to what occurs with
gap functions with nodes, where the applied
field dependence is simply  given by an overall factor of $H_a^2$, 
the angular and applied magnetic field dependences of
$m_\bot$ are now coupled, with the coupling depending on the ratio
$\Delta_s/\Delta_d$.                              
This coupling can best be described by noticing the scaling property of 
the transverse moment at any angle, which follows from 
Eq. (\ref{msid}):
\begin{equation}
m_\bot(\kappa h, \kappa \frac{\Delta_s}{\Delta_d})=
\kappa ^2 m_\bot( h,\frac{\Delta_s}{\Delta_d}),
\label{scale}
\end{equation} 
where $\kappa$ is a scaling factor.

In Fig.~\ref{msidf} we
show our results for the angular dependence of $m_\bot$ at two 
values of the dimensionless field,
$h=0.05$ and $h=0.1$. We plot $m_\bot(\psi)$ in the
range $\psi \in [0,\pi/4]$ and this can be extended from the odd parity, 
$m_\bot(\pi/4+\psi)=-m_\bot(\pi/4-\psi)$, to the range of its period, $\pi/2$.
The solid curve represents the $d$-wave result, such that its maximum at 
field $h=0.05$ 
is normalized to unity.
As in our discussion of Fig. \ref{mspf}, we refer the reader
to previous work\cite{zv}, for the unnormalized values of
this quantity.
The broken lines represent $m_\bot(\psi)$ 
for various $\Delta_s/\Delta_d$ values,  and attain their maxima at different 
angles $\psi_M$. The range of $\Delta_s/\Delta_d$ included is greater
than  that allowed by experiment, so that the scaling
(with $\kappa=2$) given by Eq.
(\ref{scale}) can be illustrated.
The threshold effect is already evident in
this Figure. The nonlinear effects are suppressed for very small
applied fields, but recover quickly and indeed the maximum value of
the plotted quantity is slightly enhanced at larger fields. This slight
enhancement arises in part from the same reasons as in the $d+s$ case,
discussed above. Although the ``$is$'' component leads to a reduction in the
phase space available, this reduction is made up for, at larger fields,
by the suppression of whichever one of the two ``jets'' is below the
angular dependent threshold, so that the region
where the contributions of the two nodes tend to cancel is diminished. 
In Eq. (\ref{msid}), for $d$ wave, we see that   
for $0 < \psi < \pi/2$ both jets contribute. In that limit 
(from the first line in Eq. (\ref{msid})) the contributions of two jets have 
opposite signs, partially cancelling each other. With the addition of a small 
$s$-wave component there are two trends: first there is a 
small reduction in the phase space available for the quasiparticle 
jets, resulting  in smaller contributions of individual jets, but 
also there will be a region where cancelation of the two jets will be 
reduced.  

We show  next 
in Fig. \ref{mxsidf} , the variation with applied field of the
maximum value of the transverse 
moment a as a function of angle, 
for various values of 
$\Delta_s/\Delta_d$. There are several interesting features in this 
Figure. The field dependence  is
no longer $\propto h^2$, except for the $\Delta_s \rightarrow 0$ limit
(pure $d$-wave,  depicted by the solid line). 
The broken lines (corresponding to several 
values of  $\Delta_s/\Delta_d$ as in the previous  Figure) display 
the existence of the  threshold
field $h_T$. The value of $h_T$ can be clearly seen from the Figure to
be $\Delta_s/\Delta_d$, as given above.  There is also, as previously 
discussed, a region of parameter space $h,\Delta_s/\Delta_d$, with a   
slightly enhanced $m_{\bot,max}$ compared to the $d$-wave case.
  
\section{conclusions}
We have shown in this paper how, at low temperature, the nonlinear magnetic 
moment can serve as a high quality bulk  probe to do ``node spectroscopy''
in HTSC's, that is, to investigate the 
position of  the nodes (or quasinodes) in pairing states with possible
mixed symmetries.  We have obtained analytic results for 
$m_\bot(\psi)$ and a variety of pairing states.
Analytic results are possible because
the low temperature nonlinear response due to 
quasiparticle excitations depends on the local properties of the  FS and 
of the energy gap near the nodes. We have examined these effects on $m_\bot$ 
comparing them to previous results\cite{ys,zv} for a pure $d$-wave and 
isotropic FS. For gaps with nodes, 
$m_\bot$ is $\propto H_a^2$ and we have concentrated on the angular dependence 
of $m_\bot(\psi)$.

In sections II and III, we have shown that even a small (real) 
admixture of an $s$ component to the dominant $d$-wave OP, would be 
detectable in $m_\bot(\psi)$. We have found that
$m_\bot(\psi)$ is closely linked to the direction of 
${\bf v}_f$ at the nodes (quasinodes) in the energy gap. 
Even for an isotropic FS, the
directions of the ${\bf v}_f$ at the nodes are no longer mutually 
orthogonal: the nodes become inequivalent and the period of 
$m_\bot(\psi)$ doubles to $\pi$. Experimentally, as can be seen 
from Fig. \ref{mspff}, it may still be convenient to examine the $\pi/2$ and 
$\pi$ Fourier components of $m_\bot(\psi)$ to determine the admixture 
of $s$-wave. We also have studied the
effect of the $a-b$ plane penetration depth 
anisotropy on the pure $d$-wave state. If experimental values  as reported for 
YBCO\cite{ybco} are used, such an anisotropy significantly alters 
the corresponding $d$-wave result with an isotropic FS.
We have included the
influence of  orthorhombic distortion on $m_\bot(\psi)$, 
for both the YBCO and BSCCO families of cuprates. These effects 
can be understood in terms of an angle $\nu_A$ reflecting the combined 
effects of an anisotropic 
FS and the admixture of $s$ component.

In Section \ref{noless} we have investigated gap functions with 
quasinodes. The concrete example of an $s+id$ gap clearly shows that $m_\bot$ 
is a very sensitive probe that allows one to distinguish nodes from 
quasinodes. To do so, one must consider both its angular and field 
dependences.  
The angular 
dependence is no longer independent of the applied field, and the field 
dependence is not $\propto H_a^2$. 
There is a characteristic threshold field $H_T=H_0 \Delta_s/\Delta_d$
required to create quasiparticles. No 
nonlinear effects are present for $H_a < H_T$. 
These distinct features of $m_\bot(\psi)$ strongly suggest that they 
are generic to the other forms of gap function with quasinodes. 

We have focused here on clean superconductors. This is because, as 
shown in the extensive discussion of Refs. \onlinecite{sv,ys}, 
impurities are unimportant, for clean available samples, except at very 
small applied fields. 
The methods that we have described in this paper can be readily extended 
to other forms of gap functions, such as those that 
attribute an explicit role to the presence of chains\cite{carbch}
in YBCO-like cuprates, or to other combinations not included here. 

Possible future extensions of this work also include incorporating
an explicit time 
dependence of the applied field, since the use of more sensitive 
experimental AC techniques to measure the nonlinear effects is
contemplated.\cite{anandpc}
\acknowledgments
We thank A. Bhattacharya and A. M. Goldman for many useful 
conversations related to the experimental work on the nonlinear magnetic 
moment. We also thank J. A. Sauls and B. P. Stojkovi\'c for discussions. 
\appendix
\section {currents}
\label{cure}
We give here the general analytic expression for ${\bf j}_{qp}{\bf 
(v)}$ at low temperature. For a superconducting  gap 
that can be approximated at the node $n$ by 
$\left|\Delta(\phi_n)\right| \approx \left|\mu \Delta_{eff, n}\phi_n \right|$, 
$n=1,2,...$, elementary integration of Eq. (\ref{jq}) yields:
\begin{eqnarray}
\label{jqap}
{\bf j}_{qp}{\bf (v)}&\approx&-2e\sum_n N_{fn} v_{fn} {\hat x}_n \mu 
\Delta_{eff,n} 
\int^{\phi_{cn}}_{-\phi_{cn}}  \frac{d \phi_n}{2 \pi} 
[\phi^2_{cn}-\phi_n^2]^{1/2} \\
&=&-\frac{e}{2}\sum_n N_{fn} {\hat x}_n v_{fn}^3 
\frac{({\bf v} \cdot {\hat x}_n)^2}{\mu \Delta_{eff,n}}, 
\qquad n=1,2,... \nonumber
\end{eqnarray}
where for various directions of the applied field,  
summation will be over different nodes and $\phi_{cn}$ is given by Eq. 
(\ref{phicds}). For YBCO, including the effect of 
FS anisotropy, we set $\Delta_{eff,n}=\Delta_{eff}$ (see section 
\ref{node}), 
and obtain that 
nodes $3,\;4$ contribute to 
 ${\bf j}_{qp \:a,b}$  if $\psi \in [\psi_1, \psi_2]$. At 
$\psi=\psi_1$ only node 3 contributes to the quasiparticle 
excitations, therefore the expressions for 
 ${\bf j}_{qp \:a,b}$ from Eq. (\ref{jqap}) with summation over nodes 
$2,\;3$ and over nodes  $3,\;4$ should coincide. Writing explicitly 
 ${\bf j}_{qp \:a}$, quadratic in $h$, in dimensionless form we 
have:
\begin{equation}
e_{1a}(\psi) L^2_a(Z)+ e_{3a}(\psi) L^2_b(Z)
=e_{2b}(\psi) L_a(Z) L_b(Z), \qquad \psi=\psi_1 
\label{dom1}
\end{equation}
which gives (we recall Eq. (\ref{alinsol})):
\begin{equation}
\tan(\psi_1(Z)-\frac{\pi}{4}) \equiv 
-\frac{\lambda_b}{\lambda_a}\tan(\frac{\pi}{4}+\nu_A)
\frac{\tanh(r_a Z)}{\tanh(r_b Z)}.
\label{om1}
\end{equation}
Unlike in the case of an isotropic FS, the angular range of the 
applied field contributing to certain nodes is a function of $Z$. 
For a slab we can simplify various expressions by taking 
$\psi_1(Z)\sim\psi_1(Z_s)\equiv \psi_1$, and this was done in Eq. (\ref{ampsi}). 
In an analogous way we obtain that $\psi_2(Z)=-\psi_1(Z)$ as well as 
the range for $\psi$ where other nodes contribute.
This approximation produces a very small discontinuity in the
results at the end points of the intervals in Eq.~\ref{mpsi}. This has
been interpolated over in Fig.~\ref{mybcof}. 
\section {nonlinear velocity for an anisotropic FS}
\label{coef}
After evaluating  ${\bf j}_{qp}$, as discussed in Appendix \ref{cure}, we can 
cast Eq. (\ref{lap}) in dimensionless form. From the perturbation 
method the nonlinear part in velocity, ${\bf N}$, satisfies Eq. 
(\ref{neqa}), which is valid for any $\psi$. The coefficients in 
Eqs. (\ref{ahomo}) and (\ref{apart}) are:
\begin{mathletters}
\label{coeffic}
\begin{eqnarray}
A_{1i}&=&-A_{2i} \tanh(r_i Z_s)-\frac{2 r_a}{r_i}P_i\frac{\sinh(2 r_a Z_s)}
{\cosh(r_i Z_s)}
-\frac{r_a+r_b}{r_i}R_i\frac{\sinh((r_a+r_b)Z_s)}{\cosh(r_i Z_s)}\nonumber \\
&-&\frac{r_a-r_b}{r_i}S_i\frac{\sinh((r_a-r_b)Z_s)}{\cosh(r_i Z_s)}
-\frac{2 r_b}{r_i}T_i\frac{\sinh(2 r_b Z_s)}{\cosh(r_i Z_s)},\\
A_{2i}&=&-P_i-S_i-T_i-U_i, \qquad i=a,b \nonumber
\end{eqnarray}
\begin{eqnarray}
P_a=-\frac{h^2 e_{1a}}{\mu^2 6r_a^4}\frac{\cos^2 \omega}{\cosh^2 (r_a 
Z_s)}, \qquad 
  P_b=-\frac{h^2 e_{1b}}{\mu^2 2r_a^2(4r_a^2-r_b^2)}
	 \frac{\cos^2\omega}{\cosh^2(r_a Z_s)},
\end{eqnarray}
\begin{eqnarray}
R_a&=&-\frac{h^2 e_{2a}}{\mu^2 2r_a 2r^2_b(2r_a+r_b)}
\frac{\cos\omega
\sin \omega}{\cosh(r_a Z_s)\cosh(r_b Z_s)}, \\
R_b&=&-\frac{h^2 e_{2b}}{\mu^2 2r_a^2r_b(r_a+2r_b)}\nonumber
\frac{\cos \omega
\sin \omega}{\cosh(r_a Z_s)\cosh(r_b Z_s)},
\end{eqnarray}
\begin{eqnarray}
S_a&=&-\frac{h^2 e_{2a}}{\mu^2 2r_ar_b^2(2r_a-r_b)}
\frac{\cos \omega
\sin \omega}{\cosh(r_a Z_s)\cosh(r_b Z_s)},
\\
S_b&=&\frac{h^2 e_{2b}}{\mu^2 2r_a^2r_b(r_a-2r_b)}
\frac{\cos \omega
\sin \omega}{\cosh(r_a Z_s)\cosh(r_b Z_s)}, \nonumber
\end{eqnarray}
\begin{eqnarray}
T_a=-\frac{h^2 e_{3a}}{\mu^2 2r_b^2(4r_b^2-r_a^2)}
\frac{\sin^2 \omega}{\cosh^2(r_b Z_s)}, \qquad
T_b=-\frac{h^2 e_{3b}}{\mu^2 6r_b^4}
\frac{\sin^2 \omega}{\cosh^2 (r_b Z_s)}, 
\end{eqnarray}
\begin{eqnarray}
U_a&=&-\frac{h^2}{\mu^2 2 r_a^2}
(\frac{e_{1a}\cos^2  \omega}{r_a^2\cosh^2 (r_aZ_s)}
+\frac{e_{3a}\sin^2  \omega}{r_b^2\cosh^2 (r_bZ_s)}),
\\		
U_b&=&-\frac{h^2}{\mu^2 2 r_b^2}
(\frac{e_{1b}\cos^2 \omega}{r_a^2\cosh^2(r_aZ_s)}
+\frac{e_{3a}\sin^2 \omega}{r_b^2\cosh^2(r_bZ_s)}), \nonumber	
\end{eqnarray}
\end{mathletters}
where $\omega\equiv \psi-\frac{\pi}{4}$ and the coefficients 
$e_{i,\:1,2,3}$ can be obtained as in the isotropic case.

%
%
\begin{figure}
\caption{Coordinates and  definitions for the $d+s$ OP calculations. 
The FS and the energy gap are shown schematically. The axes $a$ and $b$ 
are along the crystallographic directions. The orthogonal $X$ and $Y$ axes 
are along the nodal directions of the pure $d$-wave gap. The $d+s$ 
nodal directions, labeled 1,2,3,4 are shifted by an angle $\pm \nu$ 
(see Eq. (\protect\ref{nu})) 
from their $\Delta_s=0$ values. 
The applied magnetic field, ${\bf H}_a$, forms an angle $\psi$ with the 
$+Y$ direction.}
\label{fig1}
\end{figure}
\begin{figure}
\caption{ Effect of an $s$ component on the
angular dependence of the nonlinear transverse magnetic 
moment.  The quantity plotted is 
$m_\bot(\psi)$, calculated for a $d+s$ gap and an isotropic FS,
at constant field, normalized so that its maximum is 
unity for the pure $d$-wave (solid line). The broken lines are labeled by the 
 corresponding values of the quantity $\nu$  as defined in Eq.
(\protect\ref{nu}). A full period is shown.}
\label{mspf}
\end{figure}
\begin{figure}
\caption{Fourier components for the case shown in Fig. 
\protect\ref{mspf}. The results shown are the 
$\pi/2$ Fourier component (solid line)  and the $\pi$ (broken line) 
 component, 
at a constant field, as a function of angle $\nu$ ( Eq.
(\protect\ref{nu})). Normalization is taken so that the $\pi/2$  component 
is unity for a pure  $d$-wave ($\nu=0$).}
\label{mspff}
\end{figure}
\begin{figure}
\caption{Definitions of various quantities for an anisotropic 
FS with YBCO type  orthorhombicity. 
The axes $a,b$, and $X,Y$, as well as the angles $\nu$ and $\psi$, are 
defined as in Fig. \protect\ref{fig1}. For clarity, only one node (node 
2 in the scheme of Fig. \protect\ref{fig1}) is shown here and the OP is not 
depicted. The angle $\nu_A$ (at the 
node shown) is the angle between $+\hat{Y}$ and ${\bf v}_{f2}$. It has 
the same magnitude at the other three nodes. The vector ${\bf V}$ is 
the superfluid velocity.}
\label{axes}
\end{figure}
\begin{figure}
\caption{Angular dependence of $m_\bot(\psi)$, for a pure $d$-wave OP and a
FS with BSCCO type  
orthorhombicity . The quantities plotted are normalized as in Fig. 
\protect\ref{mspf}. The solid line  represents the 
isotropic case,  and the dashed line the anisotropic case
with  $\lambda_{Y}/\lambda_{X}=1.1$.}
\label{mbscof}
\end{figure}
\begin{figure}
\caption{Angular dependence of $m_\bot(\psi)$, 
for a  $d+s$ OP and YBCO type 
orthorhombicity. The solid line is the normalized reference result 
(pure $d$-wave gap and an isotropic FS). The broken lines, each labeled  by the 
corresponding value of the angle $\nu_A$ (see text and Fig. 
\protect\ref{axes}) are all obtained 
for $\lambda_a/ \lambda_b=1.5$.}
\label{mybcof}
\end{figure}
\begin{figure}
\caption{Angular dependence of $m_\bot(\psi)$, for 
various admixtures of $s$-wave component in an $s+id$ energy gap. 
Results are shown for two values of the dimensionless field, h=0.05 
(panel (a)) 
and h=0.1 (panel (b)). The normalization is taken so that the 
 pure $d$-wave (solid line) maximum is  unity at  h=0.05.
The broken lines, are labeled by  the corresponding ratio of 
$\Delta_s/\Delta_d$.
 The scaling property (with $\kappa=2$) from Eq. 
(\protect\ref{scale}), can be verified by comparing the results for h=0.05
and those for h=0.1 with the appropriate ratios $\Delta_s/\Delta_d$ increased
by a factor of two.}
\label{msidf}
\end{figure}
\begin{figure}
\caption{Field dependence of the maximum of $m_\bot(\psi)$  
 for various ratios of $\Delta_s/\Delta_d$ in an $s+id$ state. The solid line 
is the pure $d$-wave result normalized so that its value is unity for h=0.05. 
The broken lines corresponds to the same values of  
$\Delta_s/\Delta_d$ as in Fig. 7b. For each value of 
$\Delta_s/\Delta_d$ the threshold field $h_T=\Delta_s/\Delta_d$ can be 
read off from the graph. Except for $\Delta_s\equiv0$, the lines are 
not parabolic.}
\label{mxsidf}
\end{figure}
%
%
\begin{table}
\caption{Spin singlet pairing states in a $CuO_2$ plane.}
\label{table1}
\begin{tabular}{ccc}
Symmetry Class&Informal name&Representative State\\
\tableline
$A_{1g}$ &$s$ $(s^+)$    &$const.$\\
$A_{2g}$ &$g$ $(s^-)$ &$xy(x^2-y^2)$\\
$B_{1g}$ &$d_{x^2-y^2} $ &$x^2-y^2$\\
$B_{2g}$ &$d_{xy}$       &$xy$\\
\end{tabular}
\end{table}

\begin{references}
\bibitem{small} We recall that the applied field $H_a$ must not exceed 
the field of first flux entry.
\bibitem{demag} That is, e.g., for a disc shaped sample. Otherwise, the 
linear correction arising from the demagnetization tensor is easily 
included.
\bibitem{yseqn} Specifically Eq. (\ref{h0}) from Ref. \onlinecite{ys}. Our 
definition of $H_0$ differs from the one in that work by a factor of three.
\bibitem{eight} In some cases, the ``extended s-wave'' gap 
could have a multiple of eight nodes (see Ref. \protect\onlinecite{dj}).
\end{references}
\end{document}